\documentstyle[aps,preprint]{revtex}
\input psfig
\begin{document}
\draft
\title{Suppression of antiferromagnetic correlations\\ 
by quenched dipole--type impurities}
\author{V. Cherepanov, I. Ya. Korenblit, Amnon Aharony, and O. Entin-Wohlman}
\address{School of Physics and Astronomy, Raymond and Beverly Sackler Faculty 
of Exact Sciences,\\
Tel Aviv University, Tel Aviv 69978, Israel}
\date{\today}
\maketitle
\begin{abstract}
The effect of quenched random ferromagnetic bonds
on the antiferromagnetic 
correlation length, $\xi_{2D}$, 
of a two--dimensional Heisneberg model is studied, applying 
the renormalization group
method to the classical non--linear sigma model with quenched random dipole
moments. It is found that the 
antiferromagnetic long range order is destroyed for any 
non--zero concentration, $x$, of the dipolar defects, even at zero temperature.
Below a line $T\propto x$, where $T$ is the temperature, $\xi_{2D}$ is 
independent
of $T$, and decreases exponentially with $x$. At higher temperatures, it
decays exponentially with $\rho_{s}^{{\rm eff}}/T$, with an effective
stiffness constant $\rho_{s}^{{\rm eff}}$, which decreases with $x/T$. 
The results are
used to estimate the three--dimensional N\'{e}el temperature, which
decays linearly with $x$ at small concentrations, and drops precipitously at 
a critical concentration. These predictions are compared with
experiments on doped copper oxides, and are shown to reproduce successfully
some of the prominent features of the data.

\end{abstract}
FOR 2 COLUMN ACTIVATE THE LINE BELOW
%]

\section{Introduction}

Consider a Heisenberg
antiferromagnet, with a concentration $x$ of quenched random nearest neighbor
ferromagnetic (FM) bonds. These bonds compete with the antiferromagnetic (AFM)
order, and introduce frustration into the problem. If the FM exchange on
the ``impurity" bonds is sufficiently strong, then the two spins at the end 
of each such bond prefer energetically to be parallel to each other, and
perpendicular to the background AFM staggered moment. The staggered moments
on the other sites then cant in this perpendicular direction, and at large
distance this canting angle decays with distance $r$ as $1/r^{d-1}$ in
$d$ dimensions, 
similarly to the decay of magnetic moments in the presence of a magnetic 
dipole.\cite{villain}
This follows from a mapping of the low temperature equations for
the spin configuration at the minimal energy onto the Laplace equation.
At sufficiently low $d$ such impurities destroy the AFM long range order at
all finite temperatures $T$, giving rise to a spin glass like phase.
\cite{aharony}
The present paper presents a renormalization group (RG) analysis of the
AFM correlations in the presence of such quenched
randomness.
As we show, this complex random problem is exactly renormalizable
in two dimensions (2D), allowing a detailed study of the dependence
of the 2D AFM correlation length $\xi_{2D}$ on $T$ and on $x$.
This also allows an $\epsilon$-expansion in $d=2+\epsilon$ dimensions.
The three dimensional N\'eel temperature $T_N(x)$ of lamellar system is estimated
by a model of weakly coupled planes.

It is hoped that our model can describe the $T-x$ phase diagram of
various antiferromagnets with random FM bonds.
One motivation of the present study arises from its possible
relevance to the understanding
of the doping dependence of the magnetic order in the lamellar copper oxides.
Experimentally, doping such oxides leads to a rapid
decrease in both $\xi_{2D}$ and $T_{N}$.
\cite{review,keimer,shirane,saylor,cho,chou,kasegava,wakimoto}
Experiments on doped La$_2$CuO$_4$ show that
above $x=x_{c} \approx 2\%$ $\xi_{2D}$ remains finite even at zero temperature,
and there is no AFM order. 
This strong effect of the doping has been attributed
to frustration, due to strong FM exchange on
the Cu--O--Cu bonds which have localized holes due to the doping. \cite{aharony}
This frustration was also predicted to yield a magnetic spin glass phase for
$x>x_{c}$, \cite{aharony} as recently confirmed in detail
in doped La$_2$CuO$_4$. \cite{choua}
The experimental verification of this spin freezing above $x_{c}$
\cite{keimer,chou,wakimoto,choua,hayden,nieder} confirms the picture of
{\it localized}
holes. This localization is also confirmed by direct conductance measurements
at $x\leq 5\%$ and low temperature $T$. \cite{keimer,hayden}

The description of doped lamellar cuprates by quenched FM bonds requires
several assumptions, which will be discussed in detail below. In particular, at
high temperature $T$ the ``dipolar" moments which describe the dopant bonds
may fluctuate, turning this aspect of the problem into one which
requires a combined annealed and  quenched averaging.
Indeed, Glazman and Ioselevich (GI) \cite{glazman} studied this
problem in its annealed limit,
and calculated $\xi_{2D}$ to leading order in $x/T$
(See also Ref. \onlinecite{korn}.)
Although the locations ${\bf r}_{\ell}$
of the dipole--like impurities are randomly {\it quenched},
each impurity involves an effective dipole moment ${\bf m}({\bf r}_{\ell})$
which is still free to reach {\it annealed} equilibrium in the
presence of all the other dipoles. At low temperature, these
moments freeze as in a dipole glass. However, unlike the dipole
glass, where the interactions are fixed,
the interactions among the dipoles are mediated by the
AFM spin background, whose behavior also depends on
temperature, concentration and configuration of the dipole moments.
In the absence of a simple systematic way to handle such a combined
quenched--annealed problem, GI stopped their explicit calculations
at the low--$x$ expansion. Here we argue that at sufficiently low
$T$ the dipole moments, which interact via randomly quenched dipole--dipole
interactions, freeze in a random spin glassy way. \cite{binder,hertz}
At low temperatures we thus compare the experiments with our quenched theory.
The actual fitting of data from the cuprates should involve some
interpolation between the annealed and quenched limits.
As we show, the annealed and
quenched calculations coincide to leading order in $x/T$, and therefore
our quenched results supply a good interpolation over the whole range.
Other assumptions, concerning e. g. the mobility of the holes forming
the dopant bonds, will be discussed below.
In any case, our model should give a good description of the lamellar
cuprates at low $T$, and should describe other lamellar antiferromagnets
doped with FM bonds.

A second major motivation concerns the fact that, as
we show below, the quenched random dipoles are coupled to the
gradient of the order parameter, and therefore they are
equivalent to correlated random fields, whose correlations in
momentum space are proportional to the square of the momentum. Such
fields lower the critical dimensions of the random field
${\cal N}$--component spin model by 2, from 6
to 4 for the upper critical dimension, and from 4 to 2 for
the lower one. In Ref. \onlinecite{lacour} this has been
established for the limit ${\cal N}\rightarrow\infty $.
Here we show that
the lower critical dimension is shifted down from $d=4$ to $d=2$, for all
${\cal N} > 2$. 
As a result, both the temperature
and the variance of the random dipole moments (which is proportional
to the concentration, $x$) are marginal (in the RG sense) at $d=2$, allowing for
an analytical solution of the recursion relations.
This marginality of the randomness is related to the 2D infrared divergence
of the Villain canted states.\cite{villain}
Within the one--loop approximation, we 
obtain an exact expression for the exponential part of the
2D correlation length, which remains finite at all non--zero $x$.

We describe the system by the reduced Hamiltonian (i.e., the Hamiltonian
divided by the temperature $T$)
\begin{equation}
{\cal H}={\cal H}_{\rm pure}+{\cal H}_{\rm int},
\end{equation}
where ${\cal H}_{\rm pure}$ is the classical non-linear sigma
model (NL$\sigma $M) for the
pure (non--random) system, representing the long wave length
Hamiltonian related to the fluctuations of the unit vector
${\bf n}({\bf r})$ of antiferromagnetism,
\begin{equation}
{\cal H}_{\rm pure}=\frac{1}{2t}\int d{\bf r}\sum_{i,\mu}(\partial_{i}
n_{\mu})^{2},\ \ t=T/\rho_{s}.
\end{equation}
Here $i=1,..,d$ and $\mu =1,..,{\cal N}$ run over the spatial
Cartesian components and over the spin components, respectively, $\rho_s$
is the stiffness constant and
$\partial_{i}\equiv\partial /\partial x_{i}$.
This Hamiltonian is known to give an excellent description of the
undoped antiferromagnet, both theoretically \cite{CHN} and experimentally
\cite{keimer}. The success of this description results from the fact
that, although the problem involves quantum spin fluctuations,
these
can be integrated out at any finite $T$, causing only a renormalization
of $\rho_s$.

${\cal H}_{\rm int}$ is constructed \cite{glazman} to reproduce the
dipolar effects at long distances: Denoting by ${\bf a}({\bf r}_{\ell})$ the unit
vector directed along the doped bond at ${\bf r}_{\ell}$, and by
$M{\bf m}({\bf r}_{\ell})$ the corresponding dipole moment
(where ${\bf m}({\bf r}_{\ell})$ is a unit vector giving the
direction of the dipole, and $M$ is its magnitude),
\begin{equation}
{\cal H}_{\rm int}=\frac{1}{t}\int d{\bf r}\sum_{i}{\bf f}_{i}
({\bf r})\cdot\partial_{i}{\bf n},\label{hint}
\end{equation}
with
\begin{equation}
{\bf f}_{i}({\bf r})=M\sum_{\ell}\delta ({\bf r}-{\bf r}_{\ell})
a_{i}({\bf r}_{\ell}){\bf m}({\bf r}_{\ell}).\label{f}
\end{equation}
Note that since ${\bf n}$ is a unit vector, $\partial_{i}{\bf n}$ is
perpendicular to ${\bf n}$ and hence ${\cal H}_{\rm int}$ contains only
the ${\cal N}-1$ components of the ${\cal N}$--component vector
${\bf f}_{i}$ which are transverse to ${\bf n}$. However, since the
vector ${\bf n}$ varies with ${\bf r}$, all the components of ${\bf m}$
may enter at the end.
As stated, GI treated the variables ${\bf m}$ as annealed. Here we
treat all the variables  
${\bf r}_{\ell}$, ${\bf a}({\bf r}_{\ell})$, 
and ${\bf m}({\bf r}_{\ell})$ as quenched. Denoting quenched averages
by $[...]$, we write
\begin{eqnarray}
\bigl[{\bf a}({\bf r})\bigr ]&=&0, \nonumber\\
\bigl[m_{\mu}({\bf r})\bigr ]&=&0, \nonumber\\   
\bigl[a_{i}({\bf r})a_{j}({\bf r}')\bigr ]&=&\delta_{ij}
\delta ({\bf r}-{\bf r}')x/d,\nonumber\\   
\bigl[m_{\mu}({\bf r})
m_{\nu}({\bf r}')\bigr ]&=&
Q\delta_{\mu\nu}\delta ({\bf r}-{\bf r}'),
\label{quench}
\end{eqnarray}
so that
$[f_{i\mu}]=0$ and 
\begin{equation}
\bigl [f_{i\mu}({\bf r})f_{j\nu}({\bf r}')\bigr ]
=\lambda \delta_{\mu\nu}\delta_{ij}
\delta ({\bf r}-{\bf r}'),
\label{ff}
\end{equation}
with
\begin{equation}
\lambda =M^{2}Qx/d\equiv Ax,\label{lambda}
\end{equation}
where $A={\cal O}(1)$. At low $T$ one expects $Q \approx 1/{\cal N}$.

One can see now that ${\cal H}_{\rm int}$ represents random fields with quenched
correlations: Fourier transforming the variables in Eq. (\ref{hint}),
${\cal H}_{\rm int}$ can be written in the form
\begin{equation}
{\cal H}_{\rm int}=\frac{1}{(2\pi )^{d}t}\int d{\bf k}\sum_{\mu}
h_{\mu}({\bf k})n_{\mu}(-{\bf k}),
\end{equation}
with the random field ${\bf h}({\bf k})$,
\begin{equation}
h_{\mu}({\bf k})=i\sum_{j}k_{j}\int d{\bf r}f_{j\mu}({\bf r})
e^{i{\bf k}\cdot{\bf r}},
\end{equation}
which has quenched correlations
\begin{equation}
\bigl [h_{\mu}({\bf k})h_{\nu}^{\ast}({\bf k}')\bigr ]
=\lambda k^{2}\delta_{\mu\nu}\delta ({\bf k}-{\bf k}').
\end{equation}
Such correlations shift the critical dimension of the random
field Heisenberg problem down to 2. 
A heuristic way to show this follows Imry and Ma \cite{imry}
in assuming an ordered state and considering the
transverse spin fluctuations, ${\cal M}^{\perp}({\bf r})$.
In momentum space, 
${\cal M}^{\perp}({\bf q})=G^{\perp}({\bf q})h^{\perp}({\bf q})$,
with $G^{\perp}({\bf q})\sim 1/q^{2}$, and $h^{\perp}({\bf q})$ denoting the
transverse random field. Thus
\begin{eqnarray}
& &\bigl [ {\cal M}^{\perp}({\bf r}){\cal M}^{\perp}({\bf r}')\bigr ]
=\Bigl (\frac{1}{2\pi}\Bigr )^{2d}\int d^{d}qd^{d}q'\times\nonumber\\
& &G^{\perp}({\bf q})
G^{\perp}({\bf q}')
\bigl [ h^{\perp}({\bf q})h^{\perp}({\bf q}')\bigr ] e^{i({\bf q}\cdot {\bf r}
+{\bf q}' \cdot {\bf r}')}.\label{ym}
\end{eqnarray}
For 
$\bigl [ h_\mu({\bf q})h_\nu({\bf q}')\bigr ] \propto 
\delta_{\mu \nu} q^{\theta}\delta ({\bf q}+{\bf q}')$,
this integral diverges for $d<4-\theta $, implying that the assumption of
long range order is invalid. This identifies the lower critical dimension as
$d_{{\ell}}=4-\theta $, and for our case, $d_{{\ell}}=2$.
The calculations below support
this picture.
In addition to giving an exact solution for the problem at hand, we note that
the present formalism might also be used as a starting point for a double
expansion, in $\epsilon$ and in $2-\theta$, aiming at other random field
problems.

The conventional method to treat the NL$\sigma $M expands the
order parameter unit vector ${\bf n}$ about a spatially uniform ordered state.
\cite{brezin,nelson}  Using replicas to handle the quenched randomness we
have found that this approach generated local random fields, which
seem to break the symmetries of the original model. Similar problems were
found for other problems near 4D, where they required a resummation of the
perturbation expansion. \cite{fisher}  In our case, the
problems may have come either from the replicas or from the assumption
of a spatially uniform state. Such a state does not exist when there is
some local freezing of the moments in random directions (as happens
in the random case discussed here). We avoid both of these by going back
to the original RG approach by Polyakov, \cite{polyakov}  and by
treating the randomness without replicas. In 2D, this allows us to
obtain $\xi_{2D}$ for all values of $\rho_{s}x/T$.

We show that the quenched dipoles suppress the antiferromagnetic
correlations, so that the correlation length is a decreasing function of
$\rho_{s}x/T$, remaining finite at any non--zero $x$, even at $T=0$.
This implies the destruction of the antiferromagnetic long range order
in 2D for any concentration. As $x$ increases, $\xi_{2D}(T=0)$ decays
exponentially, representing the sizes
of the Imry-Ma domains for this system.

In order to compare the results of our model with experiments performed on doped
lamellar cuprates, we consider the 3D ordering of a system of weakly
coupled planes by using the relation
$\alpha\xi_{2D}^{2}\sim 1$. Here, $\alpha $ represents either the interplane
coupling, that is the relative interplane exchange,
$J_{\perp}/J\sim J_{\perp}/2\pi\rho_{s}$, or the in--plane relative spin
anisotropy (in the presence of which even an infinitesimal coupling
suffices to yield 3D ordering). Although it is not expected to give the 
correct 3D critical behavior, this procedure proved to give excellent results
for the 3D N\'eel temperature in the pure case. \cite{keimer} 
This ``mean field" procedure is also justified by an
RG argument: to linear order in $\alpha$, the RG recursion relation for
$\alpha$ is $\alpha'=e^{2\ell}\alpha$, where $e^\ell$ is the length rescale
factor. After $\ell$ iterations of the RG the effective coupling between
planes involves renormalized spins, contained in the area $e^{2\ell}$ of the
renormalized cell. A 3D behavior is expected when this effective coupling
becomes comparable to 1.
As we show in Appendix B, similar results for the phase diagram
are also found from integrating
the RG recursion relations in $d=2+\epsilon$ dimensions.
We obtain an explicit form for $T_{N}(x)$, 
and compare it in detail with available data. Our results reproduce 
prominent features of the observed phase diagram, in particular the fast decrease
of the N\'{e}el temperature with increasing $x$ and the disappearance of the
3D long range order at $x\sim 2\%$.

The outline of the paper is as follows. Section II discusses the RG procedure,
and Sec. III describes the RG recursion relations for the quenched averaging.
The 2D recursion relations are then
solved in Sec. IV, and the resulting $\xi_{2D}$ is used for estimating
the 3D phase transition line $T_{N}(x)$ in Sec. V. Section VI then
contains a discussion of the alternative annealed averaging.
The results are compared
with experiments on doped cuprates in Sec. VII, and discussed in
Sec. VIII. Details of the calculations
and extensions to $d=2+\epsilon $ are given in the Appendices.

\section{The renormalization group procedure}

Following the RG approach of Polyakov,
\cite{polyakov,berezinskii} we decompose ${\bf n}({\bf r})$ into a slowly
varying part, given by the unit vector ${\bf {\tilde{n}}}({\bf r})$,
and ${\cal N}-1$ fast variables $\phi_{\mu}({\bf r})$, such that
\begin{eqnarray}
{\bf n}({\bf r})=& &{\bf {\tilde{n}}}({\bf r})
\sqrt{1-\phi^{2}({\bf r})}+
\sum_{\mu =1}^{{\cal N}-1}\phi_{\mu}({\bf r})
{\bf e}_{\mu}({\bf r}),\nonumber\\
& &\phi^{2}({\bf r})=\sum_{\mu =1}^{{\cal N}-1}\phi_{\mu}^{2}({\bf r}).
\label{decom}
\end{eqnarray}
The unit vectors ${\bf {\tilde{n}}}({\bf r})$ and ${\bf e}_{\mu}({\bf r})$,
$\mu =1,..,{\cal N}-1$, form an orthonormal basis. The Fourier 
transform of the fast variables $\phi_{\mu}$,
\begin{equation}
\phi_{\mu}({\bf r})=(2\pi)^{-d}\int d{\bf q}e^{i{\bf q}{\bf r}}
\phi_{\mu}({\bf q}),
\end{equation}
is restricted to wave vectors ${\bf q}$ in the range $b^{-1}\leq q\leq 1$.
The upper bound is the inverse of the microscopic length (which is
measured in units of the lattice constant), and $b$ is the length rescale
factor for the renormalization procedure. 
These $q$ values are to be integrated
out. After the iteration the correlation length $\xi $ is renormalized into
$\xi /b$.

The Hamiltonian ${\cal H}$ requires the derivatives of ${\bf n}({\bf r})$.
Using the relations ${\bf {\tilde{n}}}\cdot\partial_{i}{\bf {\tilde{n}}}
=0$, ${\bf {\tilde{n}}}\cdot{\bf e}_{\mu}=0$, and
${\bf e}_{\mu}\cdot{\bf e}_{\nu}=\delta_{\mu\nu}$, we set
\begin{equation}
\partial_{i}{\bf {\tilde{n}}}=\sum_{\mu =1}^{{\cal N}-1}
B_{i}^{\mu}{\bf e}_{\mu},\ 
\partial_{i}{\bf e}_{\mu}=\sum_{\nu =1}^{{\cal N}-1}A_{i}^{\mu\nu}
{\bf e}_{\nu}-B_{i}^{\mu}{\bf {\tilde{n}}},
\end{equation}
where $A_{i}^{\nu\mu}={\bf e}_{\mu}\cdot
\partial_{i}{\bf e}_{\nu}=-A_{i}^{\mu\nu}$.
Then we find
\begin{eqnarray}
\partial_{i}{\bf n}&=&
{\bf {\tilde{n}}}\Bigl \{\partial_{i}\sqrt{1-\phi^{2}}
-\sum_{\mu =1}^{{\cal N}-1}B_{i}^{\mu}\phi_{\mu}\Bigr \}\nonumber\\
&+&\sum_{\mu =1}^{{\cal N}-1}{\bf e}_{\mu}\Bigl \{\partial_{i}
\phi_{\mu}+B_{i}^{\mu}\sqrt{1-\phi^{2}}+\sum_{\nu =1}^{{\cal N}-1}
A_{i}^{\nu\mu}\phi_{\nu}\Bigr \}.\label{dn}
\end{eqnarray}
We show in Appendix A that the functions $A_{i}^{\mu\nu}$, which
give the first derivatives of the base vectors ${\bf e}_{\mu}$,
can be eliminated by a suitable gauge transformation when one
ignores higher order derivatives. \cite{polyakov,patashinskii}
Therefore, these are omitted in the following.

In terms of the new variables, the Hamiltonian ${\cal H}$, 
to order $\phi_{\mu}^{2}$, reads
\begin{equation}
{\cal H}={\cal H}_0+{\cal H}_{1}+{\cal H}_{2}+{\cal H}_{3}+{\cal H}_4,
\end{equation}
with
\begin{equation}
{\cal H}_{0}=\frac{1}{2t}\int d{\bf r}\sum_{i\mu}\Bigl\{\bigl(B_{i}^{\mu}
({\bf r})\bigr)^{2}+(\partial_i\phi_{\mu}({\bf r})
)^{2}\Bigr\},\label{H0}\\
\end{equation}
\begin{equation}
{\cal H}_1=\frac{1}{t}\int d{\bf r}\sum_{i\mu}B_{i}^{\mu}({\bf r})
g_{i}^{\mu}({\bf r}),\label{H1}
\end{equation}
\begin{equation}
{\cal H}_2=
\frac{1}{2t}\int d{\bf r}\sum_{i\nu\mu}B_{i}^{\mu}
({\bf r})B_{i}^{\nu}({\bf r})\bigl\{\phi_{\mu}
({\bf r})\phi_{\nu}({\bf r})-\delta_{\mu\nu}\phi^{2}({\bf r})\bigr\},\label{H2}
\end{equation}
\begin{equation}
{\cal H}_{3}=-\frac{1}{t}\int d{\bf r}\sum_{i\mu}B_{i}^{\mu}({\bf r})
\Bigl \{u_{i}({\bf r})\phi_{\mu}({\bf r})
+\frac{1}{2}g_{i}^{\mu}({\bf r})\phi^{2}\Bigr \},\label{H3}
\end{equation}
and
\begin{equation}
{\cal H}_{4}=\frac{1}{t}\int d{\bf r}\sum_{i\mu}\Bigl\{g_{i}^{\mu}({\bf r})
\partial_i\phi_{\mu}({\bf r})-\frac{1}{2}u_{i}({\bf r})
\partial_i\phi_{\mu}^{2}({\bf r})\Bigr\}.\label{H4}
\end{equation}
Here we have introduced the notations
\begin{equation}
u_{i}({\bf r})=\tilde{\bf n}({\bf r}) \cdot {\bf f}_{i}({\bf r}),\ \ 
g_{i}^{\mu}({\bf r})={\bf e}_{\mu}({\bf r}) \cdot {\bf f}_{i}({\bf r}),\label{ug}
\end{equation}
for the longitudinal and transverse components, respectively, 
of ${\bf f}_i$ in the new variables.
Indeed, Eq. (\ref{H1}) represents the bare form
of ${\cal H}_{\rm int}$ in this system. In ${\cal H}_{0}$
the slow variables, $B_{i}^{\mu}$, are separated from the fast ones
$\phi_{\mu}$.
[One should notice that $(\partial_i n_{\mu})^{2}$ also yields 
the contribution $(1/t)\int d{\bf r}\sum_{i\mu}B_{i}^{\mu}\partial_i\phi_{\mu}$.
However, this term vanishes upon Fourier transforming, as 
$\phi_{\mu}$ pertains to the
large $q$ portion of the Brillouin zone, while the slow variables 
$B_{i}^{\mu}$ belong to
the small $q$ values, $q\leq b^{-1}$.]
The Hamiltonians ${\cal H}_{2}$, ${\cal H}_{3}$, and ${\cal H}_{4}$,
of order ${\cal O}(B^{2})$, ${\cal O}(B^{1})$, and ${\cal O}(B^{0})$,
respectively, will be treated in perturbation theory.

\section{Renormalization group equations}

The first step in deriving the recursion relations involves
integration over the fast variables $\phi_{\mu}$.
This requires the Green's functions,
\begin{eqnarray}
& &\langle\phi_{\mu}({\bf r})\phi_{\nu}({\bf r}')\rangle
=\delta_{\mu\nu}G({\bf r}-
{\bf r}'),\nonumber\\
& &G({\bf r})=(2\pi)^{-d}\int_{b^{-1}\leq q\leq 1} d{\bf q}
e^{i{\bf q}\cdot {\bf r}}{\hat G}(q),\ \
{\hat G}(q)=t/q^2,
\end{eqnarray}
where $\langle ..\rangle$ denotes a thermal average with ${\cal H}_0$.
As we shall see below, to
leading order in $\epsilon=d-2$ we need $G(r)$ only at strictly 2D, where
\begin{equation}
G(0)=\frac{t}{2\pi}\ln b.\label{G0}
\end{equation}
Hence $\langle\phi^{2}\rangle $ is small for $\ln b\ll 2\pi /t$.
In practice, $G(r)$ is significantly different from zero only for
$1 < r < b$, where it is approximately given by the 2D Coulomb interaction
\begin{equation}
G(r) \approx \frac{t}{2 \pi} \ln \frac{b}{r}. \label{Gr}
\end{equation}

We next turn to the perturbation expansion in ${\cal H}_{2}$,
${\cal H}_{3}$, and ${\cal H}_{4}$. The first order yields
\begin{eqnarray}
{\cal H}^{(1)}&=&\langle {\cal H}_2+{\cal H}_3\rangle=\frac{1}{2t}\int d{\bf r}
\sum_{i\mu}\Bigl(B_{i}^{\mu}({\bf r})\Bigr)^{2}(2-
{\cal N})G(0)\nonumber\\
&-&\frac{1}{2t}\int d{\bf r}\sum_{i\mu}B_{i}^{\mu}({\bf r})g_{i}^{\mu}
({\bf r})({\cal N}-1)G(0).\label{HH1}
\end{eqnarray}
The first term here represents the leading order renormalization of $1/t$, as
usual. \cite{nelson,polyakov}  The second term, which is linear in $B_i^\mu$, is
similar to the initial ${\cal H}_{\rm int}$, or to the equivalent Eq. (\ref{H1}).
In fact, this term contributes to the renormalization of the transverse
components of ${\bf f}_i$.

Higher order perturbations contain higher powers of $\phi$, which yield
higher powers of $G$ and hence of $t$. In the following, we keep only
leading powers of $t$.
Neglecting the terms involving products of two $G$'s, the second order
perturbation yields
\begin{eqnarray}
& &{\cal H}^{(2)}=-\frac{1}{2}\langle ({\cal H}_{3}+{\cal H}_{4})^{2}\rangle
=\nonumber\\
&-&\frac{1}{2t^{2}}\int d{\bf r}_{1}d{\bf r}_{2}\sum_{ij\mu}
g_{i}^{\mu}({\bf r}_{1})g_{j}^{\mu}({\bf r}_{2})
\partial_{1i}\partial_{2j}G({\bf r}_{12})\nonumber\\
&+&\frac{1}{t^{2}}\int d{\bf r}_{1}d{\bf r}_{2}\sum_{ij\mu}
B_{i}^{\mu}({\bf r}_{1})u_{i}({\bf r}_{1})g_{j}^{\mu}({\bf r}_{2})
\partial_{2j}G({\bf r}_{12})\nonumber\\
&-&\frac{1}{2t^{2}}\int d{\bf r}_{1}d{\bf r}_{2}
\sum_{ij\mu}B_{i}^{\mu}({\bf r}_{1})B_{j}^{\mu}({\bf r}_{2})
u_{i}({\bf r}_{1})u_{j}({\bf r}_{2})G({\bf r}_{12}),\nonumber\\
& &{\bf r}_{12}={\bf r}_{1}-{\bf r}_{2}.\label{HH2}
\end{eqnarray}
The first term here is $B$--independent. In principle, it gives rise to an
interaction between the dipoles: Inserting $G({\bf r})$ and the
explicit expressions for $g_{i}^{\mu}$ [Eqs. (\ref{f}) and (\ref{ug})] we 
rewrite this term in the form
\begin{equation}
{\cal H}_{\rm dd}=\frac{1}{t}
\sum_{k\ell}I_{k\ell}{\bf m}_{\perp}({\bf r}_{k})\cdot 
{\bf m}_{\perp}({\bf r}_{\ell}),\label{dd}
\end{equation}
where ${\bf m}_{\perp}({\bf r}_{\ell})$ is the component of the dipole
moment at ${\bf r}_{\ell}$ which is perpendicular to ${\bf {\tilde{ n}}}$, and
\begin{eqnarray}
I_{k\ell}&=&
\frac{1}{4\pi}M^{2}\frac{1}{r_{k\ell}^{d}}\times\nonumber\\
& &\Bigl\{
2\frac{\bigl({\bf a}({\bf r}_{k})\cdot{\bf r}_{k\ell}\bigr)
\bigl({\bf a}({\bf r}_{\ell})\cdot{\bf r}_{k\ell}\bigr)}{r_{k\ell}^{2}}
-{\bf a}({\bf r}_{k})\cdot{\bf a}({\bf r}_{\ell})\Bigr\},
\end{eqnarray}
with $r_{k\ell} < b$. 
Apart from trivial factors, this reproduces the effective
dipole--dipole interaction found in Ref. \onlinecite{glazman}.
There, Eq. (\ref{dd}) was used to integrate over the variables ${\bf m}_\perp$,
treating them as annealed variables. In the present calculation we
treat the dipoles as quenched, and therefore Eq. (\ref{dd}) simply
represents an additional constant to the energy.
We return to this point in the following.
The other two ($B$--dependent) terms in 
the second order perturbation Hamiltonian, Eq. (\ref{HH2}), will contribute
to the renormalization of the temperature and the variance of the
dipolar quenched interaction.

Finally, the third order perturbation Hamiltonian,
keeping terms up to order $B^{2}$, is
\begin{equation}
{\cal H}^{(3)}=\frac{1}{6}\langle{\cal H}_{4}^{3}\rangle
+\frac{1}{2}\Bigl (\langle{\cal H}_{4}^{2}{\cal H}_{3}\rangle
+\langle{\cal H}_{4}{\cal H}_{3}^{2}\rangle
+\langle{\cal H}_{2}{\cal H}_{4}^{2}\rangle\Big ) .\label{HH3}
\end{equation}
Integrating out the variables $\phi $, it is seen that the first
term here is independent of $B$; it contributes further to the
dipole--dipole interaction. The next term in (\ref{HH3}) 
yields an expression linear in $B$,
\begin{eqnarray}
&&\frac{1}{2}\langle {\cal H}_{4}^{2}{\cal H}_{3}\rangle =-\frac{1}{2t^{3}}
\int d{\bf r}_{1}d{\bf r}_{2}d{\bf r}_{3}\sum_{ijk}\sum_{\mu}
B_{i}^{\mu}({\bf r}_{1})\times\nonumber\\
&&\Bigl\{\sum_{\nu}g_{i}^{\mu}({\bf r}_{1})g_{j}^{\nu}({\bf r}_{2})
g_{k}^{\nu}({\bf r}_{3})\partial_{2j}\partial_{3k}
G({\bf r}_{21})G({\bf r}_{31})\nonumber\\
&&-2u_{i}({\bf r}_{1})g_{j}^{\mu}({\bf r}_{2})u_{k}({\bf r}_{3})
\partial_{2j}\partial_{3k}G({\bf r}_{23})G({\bf r}_{31})\Bigr\},
\end{eqnarray}
which again contributes to ${\cal H}_{\rm int}$,
while the last term there gives a $B^{2}$ contribution
\begin{eqnarray}
&&\frac{1}{2}\langle{\cal H}_{4}{\cal H}_{3}^{2}+
{\cal H}_{2}{\cal H}_{4}^{2}\rangle =\frac{1}{2t^{3}}
\int d{\bf r}_{1}d{\bf r}_{2}d{\bf r}_{3}\sum_{ijk}\sum_{\mu}
\times\nonumber\\
&&\Bigl\{-B_{i}^{\mu}({\bf r}_{1})B_{i}^{\mu}({\bf r}_{2})
u_{i}({\bf r}_{1})u_{j}({\bf r}_{2})u_{k}({\bf r}_{3})
\partial_{3k}G({\bf r}_{13})G({\bf r}_{23})\nonumber\\
&&+2\sum_{\nu}B_{i}^{\mu}({\bf r}_{1})B_{j}^{\nu}({\bf r}_{2})
u_{i}({\bf r}_{1})g_{j}^{\nu}({\bf r}_{2})g_{k}^{\mu}({\bf r}_{3})
\partial_{3k}G({\bf r}_{12})G({\bf r}_{23})\nonumber\\
&&+\sum_{\nu}\bigl (B_{i}^{\mu}({\bf r}_{1})B_{i}^{\nu}({\bf r}_{1})
g_{j}^{\mu}({\bf r}_{2})g_{k}^{\nu}({\bf r}_{3})-\nonumber\\
&&B_{i}^{\mu}({\bf r}_{1})B_{i}^{\mu}({\bf r}_{1})
g_{j}^{\nu}({\bf r}_{2})g_{k}^{\nu}({\bf r}_{3})\bigr )
\partial_{2j}\partial_{3k}G({\bf r}_{12})G({\bf r}_{13})\Bigr \}.
\label{rrr}
\end{eqnarray}

Except for the first term in Eq. (\ref{HH1}), all the generated terms
involve the longitudinal and transverse components of the
vecotrs ${\bf f}_{i}$, $u_{i}$ and $g_{i}^{\mu}$, Eq. (\ref{ug}), which
depend on the quenched random variables ${\bf r}_{\ell}$,
${\bf a}({\bf r}_{\ell})$,
and ${\bf m}({\bf r}_{\ell})$.
Using Eq. (\ref{ff}) we thus find 
\begin{eqnarray}
\bigl[g_{i}^{\mu}({\bf r})g_{j}^{\nu}({\bf r}')\bigr]&=&\lambda
\delta_{ij}\delta_{\mu\nu}\delta ({\bf r}-{\bf r}'),\nonumber\\
\bigl[u_{i}({\bf r})u_{j}({\bf r}')\bigr]&=&\lambda
\delta_{ij}\delta ({\bf r}-{\bf r}'),\nonumber\\
\bigl[u_{i}({\bf r})g_{j}^{\mu}({\bf r}')\bigr]&=&0.\label{av}
\end{eqnarray}

We now obtain the recursion relations of the RG. Consider first
the quenched averages of the integrated Hamiltonians
${\cal H}^{(\ell)},\ell =1,2,3$. This will constitute the renormalization
of $1/t$. Rescaling the lengths by $b^{-1}$ and the slow derivatives
$B_{i}^{\mu}({\bf q})$ by $b^{d-1}$, the new temperature
prefactor, multiplying the integral over $(B_{i}^{\mu})^{2}$, obeys the
RG equation
\begin{equation}
\frac{1}{t'}=b^{d-2}\Bigl [\frac{1}{t}+\frac{2-{\cal N}}{2\pi}
\ln b +\frac{1-{\cal N}}{2\pi}\frac{\lambda}{t}\ln b \Bigr ],
\label{rrt}
\end{equation}
which is valid to first order in $\epsilon =d-2$, $t$ and $\lambda$.
To obtain this equation,
we have used Eqs. (\ref{av}) and the relation
\begin{equation}
\sum_{j}\int d{\bf r}_{2}\bigl (\partial_{2j}G({\bf r}_{12})
\bigr )^{2}=tG(0).
\end{equation}

The terms in the Hamiltonian linear in $B_{i}^{\mu}$
remain as quenched random contributions. They renormalize
$f_{i\mu}$ and yield a renormalization of its variance $\lambda $.
To obtain the recursion relation for $\lambda $, we collect all
terms linear in $B$ and write them in the form
\begin{equation}
\frac{1}{t}\int d{\bf r}\sum_{i\mu}B_{i}^{\mu}({\bf r})
\Gamma_{i}^{\mu}({\bf r}),
\end{equation}
with
\begin{eqnarray}
&&\Gamma_{i}^{\mu}({\bf r})=g_{i}^{\mu}({\bf r})\Bigl (1-
\frac{1}{2}({\cal N}-1)G(0)\Bigr )\nonumber\\
&&+\frac{1}{t}\int d{\bf r}_{1}\sum_{j}u_{i}({\bf r})
g_{j}^{\mu}({\bf r}_{1})\partial_{1j}G({\bf r}-{\bf r}_{1})\nonumber\\
&&-\frac{1}{2t^{2}}\int d{\bf r}_{1}d{\bf r}_{2}\sum_{jk}\times\nonumber\\
&&\Bigl \{\sum_{\nu}g_{i}^{\mu}({\bf r})g_{j}^{\nu}({\bf r}_{1})
g_{k}^{\nu}({\bf r}_{2})\partial_{1j}\partial_{2k}
G({\bf r}_{1}-{\bf r})G({\bf r}_{2}-{\bf r})\nonumber\\
&& -2u_{i}({\bf r})g_{j}^{\mu}({\bf r}_{1})
u_{k}({\bf r}_{2})\partial_{1j}\partial_{2k}G({\bf r}_{1}
-{\bf r}_{2})G({\bf r}_{2}-{\bf r})\Bigr\}.
\end{eqnarray}
We then find the variance of $\Gamma_{i}^{\mu}({\bf r})$ 
by the one--loop calculation, which to leading order, 
using Eqs. (\ref{av}), yields
\begin{eqnarray}
&&\Bigl [\Gamma_{i}^{\mu}({\bf r})\Gamma_{i'}^{\mu '}({\bf r}')
\Bigr ]=\delta_{ii'}\delta_{\mu\mu '}\delta ({\bf r}-{\bf r}')
\times\nonumber\\
&&\lambda \Bigl (1-({\cal N}-1)G(0)+\frac{\lambda}{t}
(2-{\cal N})G(0)\Bigr ).
\end{eqnarray}
Hence, the recursion relation for $\lambda $ is
\begin{equation}
\Bigl (\frac{\lambda}{t^{2}}\Bigr )'=b^{d-2}\frac{\lambda}{t^{2}}
\Bigl [1 -\frac{t({\cal N}-1)+\lambda ({\cal N}-2)}
{2\pi} \ln b\Bigr ].\label{rrl}
\end{equation}

One also needs to consider the fluctuations of the random terms
around their quenched averages, as well as new terms, which were
not included in the initial ${\cal H}$, but are generated by
the renormalization procedure. However, these are irrelevant.
Let us take as an example the last term in ({\ref{HH2}),
which we may write as
\begin{eqnarray}
\frac{1}{2t}\int d{\bf r}_{1}d{\bf r}_{2}\sum_{\mu ij}
W_{ij}({\bf r}_{1}{\bf r}_{2}) B_{i}^{\mu}({\bf r}_{1})
B_{j}^{\mu}({\bf r}_{2}).
\end{eqnarray}
Physically, this term describes a random
interaction among the gradients of the order parameter ${\bf n}$, which are absent 
in the original problem.
As the ensemble average contribution of this 
interaction has been analyzed above, we
need to consider here the deviation
\begin{eqnarray}
& &\delta W_{ij}({\bf r}_{1}{\bf r}_{2})=\nonumber\\
& &-\frac{1}{t}\Bigl[u_{i}({\bf r}_{1})u_{j}({\bf r}_{2})
-\lambda\delta_{ij}\delta ({\bf r}_{1}-{\bf r}_{2})\Bigr]G({\bf r}_{12}).
\end{eqnarray}
Using Eqs. (\ref{av}), it is easy to see that
\begin{eqnarray}
& &[\delta W_{ij}({\bf r}_1{\bf r}_2)\delta W_{i'j'}
({\bf r}'_1{\bf r}'_2)]=\nonumber\\
& &\frac{\lambda^2}{t^2}\bigl[\delta_{ii'}\delta_{jj'}
\delta({\bf r}_1-{\bf r}_1')\delta({\bf r}_2-{\bf r}_2') + \nonumber\\
& &\delta_{ij'}\delta_{ji'}\delta({\bf r}_1-{\bf r}_2')
\delta({\bf r}_2-{\bf r}_1')\bigr]G({\bf r}_{12})^2.
\end{eqnarray}
Since the range of $G({\bf r})$ is of order $b$, the correlations
among these generated $W$'s are short range, and in practice the $W$'s
can be treated as uncorrelated, i.e.
\begin{eqnarray}
& &[\delta W_{ij}({\bf r}_1{\bf r}_2)\delta W_{i'j'}
({\bf r}'_1{\bf r}'_2)]=\nonumber\\
& &\Delta \bigl[\delta_{ii'}\delta_{jj'}
\delta({\bf r}_1-{\bf r}_1')\delta({\bf r}_2-{\bf r}_2') + \nonumber\\
& &\delta_{ij'}\delta_{ji'}\delta({\bf r}_1-{\bf r}_2')
\delta({\bf r}_2-{\bf r}_1'
)\bigr]\delta({\bf r}_{12}).
\end{eqnarray}
A simple power counting then
shows that $(W/t)$ scales as $b^{2d-2}$, $W$ scales as $b^{d}$ and hence
\begin{equation}
\frac{d\Delta}{d\ell}=-d\Delta +O(t,\lambda).
\end{equation}
Therefore, this generated random coupling is irrelevant.
Similar arguments apply for the variances of all the other 
random terms which are generated in the renormalization procedure.

We now follow standard procedures, and use an infinitesimal length
rescale factor $b=e^{\delta \ell}$. 
To linear order in $\epsilon =d-2$, $t$ and $\lambda$,
Eqs. (\ref{rrt}) and (\ref{rrl})
now become
\begin{eqnarray}
\frac{d}{d\ell}\frac{1}{t}&=&\epsilon\frac{1}{t}+\frac{2-{\cal N}}{2\pi}
+\frac{1-{\cal N}}{2\pi}\frac{\lambda}{t},\nonumber\\
\frac{d}{d\ell}\frac{\lambda}{t^{2}}&=&\epsilon\frac{\lambda}{t^{2}}
+\frac{1-{\cal N}}{2\pi}\frac{\lambda}{t}+\frac{2-{\cal N}}{2\pi}
\frac{\lambda^{2}}{t^{2}}.\label{drr}
\end{eqnarray}
Combining these two equations yields
\begin{eqnarray}
\frac{dt}{d\ell}&=&-\epsilon t + \frac{{\cal N}-2}{2 \pi}t^2+
\frac{{\cal N}-1}{2 \pi}t \lambda,\nonumber\\
\frac{d\lambda}{d\ell}&=&-\epsilon \lambda +\frac{{\cal N}-3}{2\pi}\lambda t
+\frac{{\cal N}}{2\pi}\lambda^2.\label{drr1}
\end{eqnarray}
As noted above, $d=2$ is the lower critical dimension for the 
NL$\sigma$M with quenched dipoles. Hence, the 2D problem is exactly 
renormalizable (as done
in the next section), and one can obtain an $\epsilon$--expansion in 
$d=2+\epsilon$ dimensions (as done in Appendix B).

\section{The correlation length of a 2D Heisenberg system}

We now proceed to calculate $\xi_{2D}$. To this end we solve Eqs. (\ref{drr1})
with the initial values $t(\ell_{0})\equiv t_{0}$, and 
$\lambda (\ell_{0})\equiv \lambda_{0}$. The parameter $\ell_0$
represents some prefacing iterations.
In the simplest case, we assume that $\ell_0=0$, and thus
that
%As we discuss below, these iterations renormalize the problem into one in
%which each renormalized cell has one impurity, i. e.
$e^{\ell_0} \equiv L_0 =1$.
%\sim x^{-1/d}$.
Other choices for $L_0$  will be discussed below.
The solution is particularly
simple for the Heisenberg system, ${\cal N}=3$. 
At 2D one finds
\begin{eqnarray}
\lambda(\ell) &=&\lambda_{0}\Bigl(1-\frac{3\lambda_{0}}{2\pi}(\ell-\ell_0) 
\Bigr)^{-1},\nonumber\\
\frac{\lambda (\ell)}{t(\ell )}&-&1=
\Bigl (\frac{\lambda (\ell )}{\lambda_{0}}\Bigr )^{1/3}\bigl (
\frac{\lambda_{0}}{t_{0}}-1\bigr ).\label{sol}
\end{eqnarray}
Both $t(\ell )$ and $\lambda (\ell )$ flow away from the fixed point
$t=\lambda =0$ as $\ell $ increases. 
>From the second of Eqs. (\ref{sol}) 
it is seen that $\lambda_{0}>t_{0}$ implies $\lambda >t$, and
{\it vice versa}.

The standard scaling relation for the correlation length is
\begin{equation}
\xi(t,\lambda)=e^{\ell }\xi(t(\ell ),\lambda (\ell )).\label{xii}
\end{equation}
The correlation length $\xi $ is obtained from the matching condition
at $\ell=\ell^\ast$,
\begin{equation}
{\rm max}\bigl(t(\ell^{\ast}),\lambda(\ell^{\ast})\bigr)=2\pi  ,\label{match}
\end{equation}
where $\ell^\ast$ is chosen so that $\xi (t(\ell^{\ast}),\lambda (\ell^{\ast}))$ 
is of the order of the renormalized lattice constant. In practice, this
implies that at $\ell=\ell^\ast$, $\xi$
is a  slowly varying function of its variables, which we denote by
$\tilde{C}(t,\lambda )$. 
The first of Eqs. (\ref{sol}) gives
\begin{equation}
\ell^{\ast}-\ell_{0}=\frac{2\pi}{3\lambda_{0}}
\Bigl (1-\frac{\lambda_{0}}{\lambda(\ell^{\ast})}\Bigr ),\label{el}
\end{equation}
where $\lambda (\ell^{\ast})$ is equal to $2\pi $ 
for $\lambda_{0}>t_{0}$, and 
%$\lambda_{0}/\lambda (\ell^{\ast})$
is given by the solution of
\begin{equation}
\Bigl (\frac{\lambda_{0}}{\lambda (\ell^{\ast})}\Bigr )^{2/3}
\Bigl [\Bigl (\frac{\lambda_{0}}{\lambda (\ell^{\ast})}
\Bigr )^{1/3}-1+\frac{\lambda_{0}}{t_{0}}\Bigr ]
=\frac{\lambda_{0}}{2\pi},\label{root}
\end{equation}
for $t_{0}>\lambda_{0}$. Equations (\ref{xii}) and (\ref{el})
give the 2D correlation length,
\begin{equation}
\xi_{2D} (t,\lambda )=L_{0}\tilde{C}(t,\lambda )\exp{\Bigl [
\frac{2\pi}{3\lambda_{0}}
\Bigl (1-\frac{\lambda_{0}}{\lambda(\ell^{\ast})}\Bigr ) 
\Bigr ]}.\label{xif}
\end{equation}

In the low temperature limit, $\lambda_{0}/t_{0} \gg 1$,
we stop iterating when $\lambda(\ell^{\ast})=2\pi$, and
consequently
\begin{equation}
\xi_{2D} \approx L_{0}\tilde{C}(t,\lambda)\exp\Bigl [\frac{2\pi}{3\lambda_{0}}
\Bigl (1-\frac{\lambda_{0}}{2\pi}\Bigr )\Bigr ].\label{xi0}
\end{equation}
It follows that $\xi_{2D}$
is finite at any finite $\lambda_0$,
even as $t_{0}$ approaches zero. This implies that at
zero temperature the long--range order in 2D Heisenberg magnets
is destroyed at any small amount of defects (as indeed predicted already
by Villain.\cite{villain})
This conclusion is also supported by the observation mentioned
above, that ${\cal H}_{\rm int}$ represents
correlated random fields. The exponential
form of Eq. (\ref{xi0}) is similar to that found for other
random field problems at the lower critical dimension. \cite{imry}
Monte Carlo simulations \cite{GM} 
also suggest that the zero
temperature correlation length is finite in 2D classical Heisenberg
magnets with frustrated bonds. When the correlation length of the 2D system remains
finite at zero temperature, it 
measures the size of the Imry--Ma domains, which is given by an exponential
form like Eq. (\ref{xi0}).

When $\lambda_0 < t_0 \ll 1$, 
we can approximate $\lambda_{0}/\lambda(\ell^{\ast})$ by
$(1-\lambda_{0}/t_{0})^{3}$ [see Eq. (\ref{root})] and obtain
\begin{equation}
\xi_{2D}=L_{0}\tilde{C}(t,\lambda ) 
\exp\Bigl(\frac{2\pi}{t_{0}}\Bigl[1-\frac{\lambda_{0}}{t_{0}}
+\frac{\lambda_0^2}{3t_0^2}\Bigr]\Bigr).\label{xi1}
\end{equation}
The exponential part may be interpreted as a renormalization of the effective
stiffness constant in the usual expression for the 2D Heisenberg model,
\begin{eqnarray}
\rho_s^{{\rm eff}}&=&\rho_s(1-y+y^2/3),\nonumber\\
y&=&\lambda_0/t_0=Ax\rho_s/T,
\label{rhos}
\end{eqnarray}
where we have used Eq. (\ref{lambda}).
To leading order in $x\rho_s/T$, this coincides with the expression 
which was obtained in Ref. \onlinecite{glazman} for an annealed system of dipoles.
Indeed, up to the lowest
order in $\lambda/t$ there is no difference
between quenched and annealed averaging.

Finally, we discuss the pre-exponential factor in the expressions
for the corrletaion length. For $\lambda_0 \ll t_0$,
the prefactor $\tilde{C}(t,\lambda) \approx \tilde{C}(t_0,0)$ is known:
The two--loop\cite{CHN} and three--loop\cite{HN} calculations,
based on the quantum NL$\sigma$M, show that
\begin{equation}
\tilde{C}(t_0,0) = {e\over 8}{c\over 2\pi \rho_s}\left(1- {t_0\over 4\pi}\right),
\label{pref}
\end{equation}
where $c$ is the spin--wave velocity. 

At low tempreatures, we need the concentration dependence 
of the pre-exponential factor.
This results from higher order loops:
At 2D, $t=0$ and ${\cal N}=3$, the generalized recursion relation for $\lambda$
has the generic form
\begin{equation}
\frac{d\lambda}{d\ell}=\beta_2 \lambda^2 + \beta_3 \lambda^3,\label{2loop}
\end{equation}
with $\beta_2=3/(2\pi)$ and with $\beta_3$ of order $\beta_2^2$.
The solution for this equation reads
\begin{eqnarray}
e^{\ell-\ell_0}&=&\Bigl(\frac{\lambda_0(\beta_2+\beta_3\lambda(\ell))}
{\lambda(\ell)(\beta_2+\beta_3\lambda_0)}
\Bigr)^{\beta_3/\beta_2^2}\times\nonumber\\ 
& &\exp \Bigl[\frac{1}{\beta_2}\Bigl(\frac{1}{\lambda_0}-\frac{1}
{\lambda(\ell)}\Bigr)\Bigr],\label{solve}
\end{eqnarray}
and therefore, at $\ell=\ell^\ast$, where $\lambda(\ell^\ast)=2\pi$,
we have
$e^{\ell^\ast} \sim L_{0}x^{\beta_3/\beta_2^2} \exp[2\pi/(3 \lambda_0)]$.
Consequently, $\tilde{C}(t,\lambda )\approx
\tilde{C}(0,\lambda) \approx C_0 \lambda^\omega $, 
with $\omega=\beta_3/\beta_2^2$.

Within the approximations leading to Eqs. (\ref{xi0}) and (\ref{xi1}),
we note that the expressions in the exponentials and their first derivatives
are continuous at $\lambda_0=t_0$, up to terms of order
${\cal O}(\lambda_{0}/2\pi )$. In comparing our results with the experiment, 
we shall use these asymptotic expressions all the way to the 
line $\lambda_0=t_0$.

\section{The phase boundary for weakly coupled planes}

The 3D transition temperature $T_{N}(x)$, as function of the
defect concentration $x$, of a system consisting of weakly
coupled planes may be deduced from the relation
\begin{equation}
\alpha \xi_{2D}^{2}(t_{N},\lambda ) \sim 1.\label{3d}
\end{equation}
[Note that $\lambda $ is proportional to $x$, 
{\it cf} Eq. (\ref{lambda}).]  
The parameter $\alpha $ can be generated by an
interplane exchange, $J_{\perp}/J\sim J_{\perp}/2\pi\rho_{s}$,
or some in--plane spin anisotropy. 
As stated in the Introduction, this procedure gives excellent
estimates for $T_N$.\cite{keimer}

To obtain the critical line $T_{N}(x)$ we proceed as follows.
Using (\ref{xif}) in the relation (\ref{3d}) we find
\begin{equation}
1-\frac{\lambda_{0}}{\lambda (\ell^{\ast})}=
\frac{3\lambda_{0}}{4\pi}\ln\bigl [\alpha\bigl (L_{0}
\tilde{C}(t,\lambda )\bigr )^{2}\bigr ]^{-1}. \label{3d1}
\end{equation}
At low temperature, i.e., for $\lambda_0 > t_0$, 
we have $\lambda (\ell^{\ast})=2\pi $ and therefore
Eq. (\ref{3d1}) is almost independent of $t$. It thus yields
a critical value for the initial value of 
the variance, $\lambda_{c}$, and hence
a critical concentration, $x_{c}$, above which there is no antiferromagnetic
long--range order at any temperature
\begin{equation}
\lambda_{c}=\frac{2\pi}{1+\frac{3}{2}
\ln\bigl [\alpha\bigl (L_{0}
\tilde{C}(0,\lambda_{c} )\bigr )^{2}\bigr ]^{-1}},\label{lc}
\end{equation}
with $\lambda_{c}=Ax_{c}$.
In fact, $\xi_{2D} $ is expected to be practically independent of temperature
(except for the very weak dependence of the prefactor $\tilde{C}$)
for a range of values, $\lambda_{0}>t_{0}$, as given in Eq. (\ref{xi0}).
Therefore, the critical line $T_N(x)$ is expected to be practically
vertical for $t_N(x) < \lambda_{0}=Ax$. Below this line, one
might expect some range in which spin glass and antiferromagnetism 
co--exist, down
to a Gabay--Toulouse like line.\cite{hertz}  To obtain this region one
would need to also consider the 3D boundaries of the spin glass phase, 
and this is beyond the scope of the present paper. In any case, the
AFM ordering will persist up to the line $x=x_{c}$.

At smaller defect concentrations, or at higher temperatures,
i.e., for $2\pi \gg t_{0}>\lambda_{0}$,
Eqs. (\ref{root}) and (\ref{3d1}) give
\begin{eqnarray}
1&-&\frac{\lambda_{0}}{t_{N}(x)}=
\Bigl\{1-\frac{3\lambda_{0}}{t_{N}(0)}\Bigr\}^{1/3},\nonumber\\
\frac{1}{t_{N}(0)}&=&\frac{1}{4\pi}\ln\bigl [\alpha\bigl (L_{0}
\tilde{C}(t_{N}(0),0)\bigr )^{2}\bigr ]^{-1},\label{tnh}
\end{eqnarray}
where $t_{N}(0)$ is the N\'eel temperature of the pure antiferromagnet. 
The lowest order of this expression agrees with the 
results of Ref. {\onlinecite{glazman}, obtained by an annealed average.

\section{Annealed averaging}

As stated above, the cuprates require some mixed annealed--quenched
averaging. We start be reviewing a simple version of GI's
analysis.\cite{glazman}
In their approach, 
${\bf r}_{\ell}$ and ${\bf a}({\bf r}_{\ell})$ are treated as quenched
variables, with averages given by Eq. (\ref{quench}), while
${\bf m}({\bf r}_{\ell})$
is treated as annealed.   
Thus, Eq. (\ref{ff}) is replaced by
\begin{equation}
[f_{i\mu}({\bf r}_1)f_{j\nu}({\bf r}_2)]=\delta_{ij}\delta({\bf r}_{12})
\Lambda_{\mu\nu}({\bf r}_1),
\end{equation}
with
\begin{equation}
\Lambda_{\mu\nu}({\bf r})=
m_\mu({\bf r})m_\nu({\bf r})
M^2 x/d.
\end{equation}

Initially, ${\cal H}$ contains no interactions among the dipole moments
$\{{\bf m}({\bf r})\}$. However, the RG iterations generate the dipole--dipole
interaction, as given by Eq. (\ref{dd}). This interaction is mediated via
the canted background AFM moments.
Treating this interaction as a 
small perturbation, to lowest order,
we can next integrate the dipole moments out of the
partition function, using the annealed averaging
$\langle m_\mu({\bf r}_1)m_\nu({\bf r}_2) \rangle = \delta_{\mu \nu} \delta({\bf r}_{12})/{\cal N}$, so that
\begin{equation}
\langle \Lambda_{\mu \nu} \rangle =\delta_{\mu \nu} \Lambda
\equiv \delta_{\mu \nu}M^2x/(d {\cal N}).
\end{equation}
Note that to this leading order, $\Lambda=\lambda$! GI wrote down a more
general form for this thermal average, involving the susceptibility which
results from the quadratic coupling in ${\cal H}_{\rm dd}$. This reduces
to the above expression at lowest order.

Up to Eq. (\ref{rrr}), we have performed no averaging. In the annealed
case, the recursion relation are derived from the same equations, using
the averages as listed above. The resulting recursion relation for
$1/t$ turns out to be the same as Eq. (\ref{rrt}), with $\lambda$
replaced by $\Lambda$. In contrast, the averaging over ${\bf m}$
gives no contributions to the renormalization of ${\bf f}_i$, since all
the generated terms which are linear in $B^\mu_i$ involve odd powers of the
$u_i$'s and the $g^\mu_i$'s, and therefore odd powers of the $m_\mu$'s.
All of these vanish upon the annealed averaging over the $m_\mu$'s.
Thus, we end up
with 
\begin{equation}
\Bigl (\frac{\Lambda}{t^2} \Bigr )' = b^{d-2}\Bigl (\frac{\Lambda}{t^2} \Bigr ).
\label{ann}
\end{equation}

At 2D, Eq. (\ref{ann}) implies that $\Lambda$ is unrenormalized.
The solution of the recursion relation for $t$ then yields
\begin{equation}
\ell=\frac{\pi}{\Lambda}\ln \frac{1+2 \Lambda/t(0)}
{1+2 \Lambda/t(\ell)}.
\end{equation}
Assuming that $\Lambda \ll 2\pi$, and integrating up to $t(\ell^\ast)=2\pi$,
yields 
\begin{equation}
\ell^\ast \approx \frac{2\pi}{t(0)}\Bigl(1-\frac{\Lambda}{t(0)}\Bigr),
\label{ann1}
\end{equation}
which agrees to leading order with the quenched result (\ref{xi1}).

This annealed averaging is legitimate as long as the renormalized
distance beteen the impurities remains large, so that the dipole--dipole
interaction which is generated during the iterations remains small.
Note that the range of ${\cal H}_{\rm dd}$ is $b=e^{\ell^\ast}$.
Thus, if $e^{\ell^\ast}$ is small compared to the inter-impurity
distance $x^{-1/d}$, then we can still treat the dipoles as independent
degrees of freedom, ignore the interaction between them and continue the
above annealed calculation.
However, if $e^{\ell^\ast}$ becomes larger than (but of the order of)
$x^{-1/d}$ then each renormalized cell contains more than one impurity,
and the interaction between them comes into play.
Since this interaction decays as $1/r^d$, the
dipoles behave like a spin glass which is at its lower critical dimension.
\cite{hertz}
Since the dipole--dipole energy is of order $E_{\rm dd}=\lambda \rho_s$,
we expect the dipole moments to develop spin--glassy correlations,
with a correlation length $\xi_{\rm sg}$
which grows exponentially in $E_{\rm dd}/T$.
Since we showed that $\xi_{2D}$ remains finite at all $T$,
we expect that for $T \ll E_{\rm dd}$ one has $\xi_{\rm sg} \gg \xi_{2D}$,
the moments inside a renormalized cell (at distances smaller than $\xi_{2D}$)
freeze randomly,
with the effective Edwards-Anderson order parameter $Q$, and we can
switch to our quenched calculation.
We thus choose to perform a prefacing annealed renormalization, up to
$e^{\ell_0}=L_0 \sim x^{-1/d}$.
For larger $\ell$ we assume that the dipole moments are frozen at low $T$,
and we switch to the quenched analysis of Secs. IV and V.

\section{Comparison with experiments on doped cuprates}

Given the discussion in the previous section, we adopt the
following strategy: We start by fitting $A \equiv \lambda/x$ 
from the $t$-dependence
of $\xi_{2D}$ at high $T$ and small $x$. As stated, this dependence
is the same for both types of averages.
Given $A$, we next fit the $x$ dependence of $\xi_{2D}$ in the limit
of very low $T$, when $\xi_{2D}$ is $T$-independent. This behavior
should certainly be described by our quenched theory. The fit determines
the pre-exponential factors.
Finally, we use the results (without any further adjustments)
to calculate the phase diagram, $T_N(x)$.

We begin with the temperature dependence of the correlation length at
low concentrations, $x<x_c$.
Data taken on a sample of
La$_2$CuO$_{4+\delta}$, with $T_N= 90$K, \cite{keimer}  show a
practically linear
dependence of $(t/2\pi)\ln (\xi/C)$ on
$1/t$, in agreement with both our quenched
result (\ref{xi1}) and our annealed result (\ref{ann1}). For the coefficient $C$
in this fit we used $C=1.92\AA$, derived from Eq. (\ref{pref}) with
$\rho_s=24$meV (See Ref. \onlinecite{keimer}) and $L_0=1$. The slope is
$\Lambda=\lambda_{0}= 0.29(1)$ (see Fig. 1, and also Ref. \onlinecite{korn}).
To estimate the value of $x$ for this sample, we follow Refs.
\onlinecite{keimer} and \onlinecite{chen}
and approximate the line $T_N(x)$ by the straight line
$T_N(x) \approx 325 - 16250 x$, which extrapolates to $x=0.02$ as $T_N$ is
extrapolated to zero. Using this approximation, we find that $T_N=90$K
at $x=0.0145$. Thus, $A=\lambda/x \approx 20$. Although we have some
problems with this linear extrapolation (see below), the value of $x$
cannot be larger than $x_c \approx 0.02$, so the uncertainty in $A$ is
not more than $30\%$.
Furthermore, although $A$ might have a weak dependence on $T$ and on $x$,
this could most probably be absorbed in this error estimate.
We thus use this estimate $A=20$ in everything that follows.

Keimer {\it et al.}\cite{keimer}  measured the temperature dependence of
the correlation length for three magnetically disordered
samples of La$_{2-x}$Sr$_{x}$CuO$_4$, with $x \approx $
0.02, 0.03, and 0.04.
The error in $x$ is less than $\sim 0.005$. It is believed
\cite{keimer} that the hole concentration is about that of the
Sr ions. It was found that at
low temperatures $\xi$ does not depend on $T$, and
falls with the increase of $x$ faster
than a power law.
The $\xi (T=0)$ data is depicted 
in Fig. 2, together with the value of $\xi$ cited in Ref. \onlinecite{hayden}
for La$_{1.95}$Ba$_{0.05}$CuO$_4$.
The figure also shows the theoretical values of $\xi $,
calculated from Eq. (\ref{xi0}), with
$L_0 \tilde{C}(0,\lambda_0)=C_0 \lambda_{0}^{\omega'}$, 
$\lambda_{0}=20x$, $C_0=2.8\AA$ and $\omega'=0.8$.
(In 2D, we now have $\omega'=\omega-1/2$, from the $x$ dependence of $L_0$).
We also reproduce in Fig. 2 the numerical results obtained in
Ref. \onlinecite{GM}. These authors computed the effect of holes
localized on the oxygen atoms in the CuO$_{2}$ plane. Their results,
marked by `$+$' in Fig. 2, are in agreement with our calculation.

We mention in passing that the correlation length at low temperature was
also measured for $x > 0.05$.\cite{gsbc}  These data were
not included in our comparison, mainly because for these concentrations
the holes are probably mobile.

We next turn to the temperature dependence of $\xi$ at $x>x_c$.
Figure 3 exhibits the calculated temperature dependence of $\xi $,
for several concentrations $x$. The curves were found from Eq. (\ref{xi1})
with $C=L_0 \tilde{C}=1.92 \AA$  ($\lambda_{0}<t_{0}$, as in Fig. 1
discussed above), and from Eq. 
(\ref{xi0}), with $C(x)= 2.8 \lambda_{0}^{0.8}~\AA$ ($\lambda_{0}>t_{0}$).
We have also used $2\pi \rho_s$ = 150 meV. The theoretical lines
are for
$x$= 0.0225, 0.029 and 0.036 (instead of the experimental values 
0.02, 0.03, and 0.04
given by Keimer {\it et al.}). The values chosen 
are within the experimental error. \cite{keimer}
Since the prefactors
used in our fits differ in the limits $\lambda_{0}\ll t_{0}$
and $\lambda_{0}\gg t_{0}$, the
high-- and low--temperature portions of the plots do not match at
$\lambda_{0}=t_{0}$.
There the two segments are connected by  dotted lines. The dashed lines in 
Fig. 3 represent the heuristic expression
\begin{equation}
\xi^{-1}(T,x) =\xi^{-1}(0,x) + \xi^{-1}(T,0),
\end{equation}
 used in Ref. \onlinecite{keimer}
to fit their data. It seems that the heuristic expression
works as well; however, it seems that it has no theoretical basis.

The theoretical curves in Fig. 3 agree with the experimental 
results at temperatures
lower than $\sim 350-400$K. At higher temperatures the calculated values of
$\xi$ are smaller than the measured ones. Perhaps, at such high
temperatures thermal fluctuations come into play, causing a
decrease of $Q$ from $1/{\cal N}$ to lower values.
It is also possible that at high $T$ the holes are more mobile, so that
one should average over more than one bond per hole, thus reducing the effective
dipole moment. 
At temperatures lower than $200- 250$ K the correlation length
does not depend on the temperature up to exponentially small
terms, of the order of $\xi(0,T)^{-1}$ .\cite{keimer}  This nontrivial
property of  the correlation length is reproduced well by our calculation.

Given the  above values for  $C_0$ and $\omega$, and the value 
$\alpha = 10^{-4}$
from Ref. \onlinecite{keimer}, we solved Eq. (\ref{lc})
and found
the critical value of $\lambda$ to be  $\lambda_c= 0.366$. With $A=20$,
the critical concentration, $x_c$, for the disappearance of the
long--range order at $T=0$ is found to be $x_c= 0.0183$,
in very good agreement with
the experimental value $x_c= 0.0175$ from Ref.\onlinecite{shirane}
(but in disagreement with Ref. \onlinecite{kasegava}, which gives
$x_c=.027$).

Using  the above parameters, i.e. $A=20$, $\alpha = 10^{-4}$,  and
approximating the prefactor by a constant, $ C(t,x) \approx C(0,x_c)
= C(0,0.0183)=
1.26 \AA$, Fig. 4 shows the theoretical
concentration dependence  of $T_N$, calculated
from Eqs. (\ref{tnh}) for $t_{0}>\lambda_{0}$
and (\ref{lc}) for $\lambda_{0}>t_{0}$, with no further fitting of the parameters.
(The line $t_{0}=\lambda_{0}$ is also depicted in the figure.)
At high temperatures  the prefactor is larger, $1.9 \AA$.
However, the effect of this difference on $T_N$ is small,
since $T_N$ depends on the prefactor only logarithmically.
At small concentrations, our theoretical  $T_N(x)$ decreases with
the increase of $x$ linearly
with the rate 55 K/\% (based on the value of $A$ as determined from the
$\xi$ data). At $x=x_c$, $T_N(x_c)$ is roughly equal to $T_N(0)/3$.
Then $T_N$ abruptly falls to zero, as our calculation finds that
the correlation length is independent of the temperature for
$t_{0}<\lambda_{0}$.

Figure 4 includes the results of several experiments.
It is seen that those of  Ref. \onlinecite{chen} seem to disagree with our phase
diagram: the data fall linearly with a slope of 162 K$/\%$,
(larger by about a factor of 3 than our theoretical value, which was
extracted from the data for $\xi$), extrapolating to $x=.02$ without the jump in
 $T_N$.
However, we should note that
Chen {\it et al.}\cite{chen} determined $x$ for their O--doped samples
from Hall effect data. In Ref. \onlinecite{chen2} $T_N$ and the Hall
density of holes were measured for a Sr doped sample. The Sr concentration
in the melt was 0.0022, while the Hall measurements
gave a smaller hole density, 0.0016.
Figure 4 shows that in this case
the experimental point (with $x=0.0022$) is closer to the theoretical curve than
the data from Ref. \onlinecite{chen}.
Saylor and Hohenemser\cite{saylor}
measured $T_N(x)$ in Sr--doped samples of lanthanum cuprate.
Although their  $T_N(0)=317$ K  was somewhat less than in the
best samples of Ref. \onlinecite{keimer},
their $T_N(x)$ decreased linearly
with the rate 90 K/\% till $x=0.015$. In the region between $x=$0.015 and
0.018, $T_N$ fell from $\approx 180$ K to 20 K. This behavior of
$T_N(x)$ is close to our phase diagram, Fig. 4.

Very recently, Wakimoto {\it et al.}\cite{wakimoto} measured
the phase diagram of oxygen doped La$_{1.95}$Bi$_{0.05}$CuO$_4$, and their
$T_N(x)$ agrees qualitatively with our theory: it falls almost linearly
down to $T_N(0.012) \approx 160$K, and then drops sharply towards
$x_c \approx 0.015$. Both this small value of $x_c$ and the low--$T$
value of the correlation length near $x_c$, as measured 
in Ref. \onlinecite{wakimoto}, are consistent with our calculations,
 with $A \approx 30$.

\section{Discussion}

Most of this paper was devoted to a detailed description of the theory for
the effects of FM bonds on AFM correlations in doped antiferromagnets.
We hope that this will stimulate some material research onto such systems.
It is also hoped that this exact solutions could be used to study more
systems with correlated random fields.

In the previous section we discussed some fits of our theory (with both annealed
and quenched averaging) to data from doped lamellar oxides. As stated in
the Introduction, these fits are based on various assumptions, and thus
deserve some further discussion. Hopefully, the reader is already convinced
that our fits give a nice interpolation between the annealed and quenched
calculations, for high and low temperatures respectively. We now discuss
possible problems in using our model of localized holes for these systems.

First of all, we note that experiments on both O--doped
and Sr--doped La$_2$CuO$_4$ indicate a localization length of the hole of order
of two lattice constants.\cite{chen,chen2}  Furthermore, for Sr--doping,
the hole moves around the center of a Cu--plaquette, and this might imply a 
quadrupolar behavior,
rather than a dipolar one, for the spin canting.\cite{korn}  However, 
there are arguments in the literature, \cite{SS} 
claiming that holes which move around such a center
create Skyrmion--type distortions
of the AFM background, decaying as $1/r$. Thus, our calculation may
remain qualitatively correct, although we might have to use a different
expression for our parameter $A$.
We might also add that to linear order in $x$ one effectively considers
only single impurities, and therefore mobile and static holes will have
the same (classical) effects.
Since $A$ is fitted from the experiments anyway, it could result from any of 
the above cases.

Another alternative is to abandon the localized picture altogether.
The effects of {\it mobile} holes on the AFM correlations were considered
in many papers.
 One approach \cite{chubukov}
concluded that mobile  holes drive the system to the quantum disordered
state. 
Indeed, we solved
the {\it classical} NL$\sigma $M, and thus ignored quantum fluctuations.
However, for the undoped systems the classical expressions described
the exponential factor of the correlation length rather well.\cite{CHN,keimer}
We have checked that integration of the quantum fluctuations,
similar to that done in Ref. \onlinecite{CHN}, only renormalizes the
initial parameters of the effective classical model also in the random
case.

Another approach \cite{hone} proposed that the AFM order is
destroyed by the segregation
 of holes into striped phases.
In view of the experiments metioned above, which show that the holes are
localized (and the conductivity obeys the variable range hopping law)
at low $T$ and low $x$, we
believe that these alternative
approaches should be used only for $x > 5\%$, or at higher
$T$. Indeed, to our knowledge there is little experimental
evidence that stripes exist at $x< 5\%$, or that quantum fluctuations
were observed at low $T$ in that range.
However, we cannot rule out some crossover from our localized to the mobile
behavior
at the metal-insulator transition (around $x=5\%$ and at higher $T$).
In any case, it would be nice to find more experimental tests that
would distinguish between the various scenarios for the cuprates.

\acknowledgements
We have benefitted from many discussions with R. J. Birgeneau , A. B.
Harris, A. S. Ioselevich, M. A. Kastner, D. E. Khmel'nitskii, V. V.
Lebedev, T. Nattermann and M. Schwartz.
This project has been supported by a grant from the U. S.--Israel 
Binational Science Foundation (BSF).

\appendix

\section{Gauge transformation of the fields $\phi_{\mu}$}

In order to eliminate the functions $A_{i}^{\mu\nu}({\bf r})$ from the expression
for the derivatives $\partial_{i}{\bf {\tilde{n}}}$ [Eq. (\ref{dn})] we
introduce the transformation
\begin{equation}
\phi_{\mu}=\sum_{\mu_{1}}T_{\mu\mu_{1}}\tilde{\phi}_{\mu_{1}},\label{A1}
\end{equation}
where the coefficients $T_{\mu\mu_{1}}$ are determined by
\begin{equation}
\partial_{i}T_{\mu\mu_{1}}=\sum_{\mu_{2}}
A_{i}^{\mu\mu_{2}}T_{\mu_{2}\mu_{1}},\label{A2}
\end{equation}
with
\begin{equation}
\sum_{\mu}T_{\mu\mu_{1}}T_{\mu\mu_{2}}=\delta_{\mu_{1}\mu_{2}},\ \ 
T_{\mu\mu_{1}}=T_{\mu_{1}\mu}.\label{A3}
\end{equation}
Note that the sums run from 1 to ${\cal N}-1$. Assuming that
$A_{i}^{\mu\mu_{1}}$ is independent of $x_{i}$, the solution of
Eqs. (\ref{A2}) and (\ref{A3}) reads
$T_{\mu\mu_{1}}=T_{\mu\mu_{1}}^0 \exp(-i \kappa x_i)$, where 
$\kappa$ is a real eigenvalue of the Hermitian matrix
$i A_i$. Deviations from this approximation  
naturally involve higher derivatives of
$A_{i}^{\mu \mu_{1}}$ which are related to
higher powers of the gradients in ${\cal H}$. These are
strongly irrelevant.

Inserting the above relations into Eq. (\ref{dn}) we obtain
\begin{eqnarray}
\partial_i{\bf n}&=&{\bf {\tilde{n}}}
\Bigl[\partial_{i}\sqrt{1-\tilde{\phi }^{2}}-
\sum_{\mu}\tilde{\phi }_{\mu}
\sum_{\nu}T_{\nu\mu}B_{i}^{\nu}\Bigr]\nonumber\\
&+&\sum_{\mu}\Bigl[B_{i}^{\mu}{\bf e}_{\mu}\sqrt{1-\tilde{\phi }^{2}}
+\partial_i\tilde{\phi}_{\mu}
\sum_{\nu}T_{\nu\mu}{\bf e}_{\nu}\Bigr].\label{A4}
\end{eqnarray}
Therefore, defining
\begin{equation}
\tilde{{\bf e}}_{\mu}=\sum_{\nu}{\bf e}_{\nu}T_{\nu\mu},\ \ 
\tilde{B}_{i}^{\mu}=\sum_{\nu}B_{i}^{\nu}T_{\nu\mu},\label{A5}
\end{equation}
Eq. (\ref{A4}) takes the form of Eq. (\ref{dn}) without the 
$A_{i}^{\nu\mu}$ terms,
with $\phi_{\mu}$, ${\bf e}_{\mu}$ and $B_{i}^{\mu}$ replaced by
$\tilde{\phi}_{\mu}$, $\tilde{{\bf e}}_{\mu}$ and $\tilde{B}_{i}^{\mu}$,
respectively. For brevity, we have omitted in the subsequent derivations
the superscript $\tilde{ }$. It is seen that the gauge transformation
can be regarded as a rotation of the base vectors ${\bf e}_{\mu}$, and
reflects the arbitrariness of their choice.\cite{polyakov}

\section{The phase boundary at 2$+\epsilon $ dimensions}

For completeness, we
present here the solution of the recursion relations (\ref{drr1}) in
$d=2+\epsilon$ dimensions, and obtain the critical line
$t_{N}(x)$, where $x$ is the defect concentration. 

The recursion relations (\ref{drr1}) 
have four fixed points in the $[\lambda ,t]$ plane:
$[0,0]$, $[0,2\pi\epsilon /({\cal N}-2)]$,
$[2\pi\epsilon /{\cal N},0]$,
and $[2\pi\epsilon /(2{\cal N}-3),2\pi\epsilon /(2{\cal N}-3)]$.
For ${\cal N} > 3/2$, the first point is stable, the last one 
is doubly unstable and the other
two are unstable in one direction and stable in the other.

The solution of Eqs. (\ref{drr1}) for $\ell>\ell_0$
and Heisenberg spins (${\cal N}=3$) reads
\begin{eqnarray}
\lambda (\ell )=\lambda_{0}\Bigl [
e^{\epsilon (\ell -\ell_{0})}&-&\frac{3\lambda_{0}}{2\pi\epsilon}
\bigl (e^{\epsilon (\ell -\ell_{0})}-1\bigr )\Bigr ]^{-1},\nonumber\\
\frac{1-\lambda (\ell )/t(\ell )}
{1-\lambda_{0}/t_{0}}&=&
\Bigl [\frac{\lambda (\ell )-2\pi\epsilon /3}
{\lambda_{0}-2\pi\epsilon /3}\Bigr ]^{1/3}.\label{pit}
\end{eqnarray}

The phase boundary is identified as the line which separates 
the flow to the 
origin (in the ordered phase) from the flow to infinity 
(in the disordered phase).
It is easy to check that for $t_0>\lambda_0$, the initial point
$[\lambda_0,t_0]$ will flow to the ``pure" fixed point $[0,2\pi\epsilon]$
when
\begin{eqnarray}
1&-&\frac{\lambda_{0}}{t_{N}(x)}=\Bigl (1-\frac{3\lambda_{0}}
{2\pi\epsilon}\Bigr )^{1/3},\nonumber\\
&&0\leq\lambda_{0}\leq\lambda_c =\frac{2\pi\epsilon}{3},\label{g}
\end{eqnarray}
where the second of Eqs. (\ref{pit}) has been used.
On the other hand, 
using again that equation, we find that for
$t_0<\lambda_0$ the solution flows to the ``random"
fixed point, $[2\pi\epsilon/3,0]$, provided that
\begin{equation}
\lambda_0 \equiv\lambda_{c}=\frac{ 2 \pi \epsilon}{3}, \;\;\;
 \lambda_{c}> t_0 > 0.\label{k}
\end{equation}
The two portions of the critical line, Eqs. (\ref{g})
and (\ref{k}), are separated by 
the multicritical point at $[2\pi\epsilon/3,2\pi\epsilon/3]$.
It is interesting to note that also here, like the behavior found for
$t_{N}(x)$ for weakly coupled planes, there is a vertical 
section of the phase boundary.

We thus conclude that for $t_0>\lambda_0$
the randomness is irrelevant, and the ``pure" critical
behavior (which has a correlation length exponent $\nu=1/\epsilon$)
remains stable. However, finite values
of $\lambda$ yield a correction to this behavior, with exponent 
$-\epsilon$ (or, more generally, $-\epsilon/({\cal N}-2)$).
The correlation length is obtained from Eq.
(\ref{xii}), using the matching conditions (\ref{match}).
One finds
\begin{equation}
\xi\sim\exp (\ell^{\ast})=L_{0}\Bigl [\frac{\lambda^{-1} 
(\ell^{\ast})-\lambda_{c}^{-1}}{\lambda_{0}^{-1}-\lambda_{c}^{-1}}
\Bigr ]^{1/\epsilon}.
\end{equation}
For $t_{0}>\lambda_{0}$ the iterations are stopped at
$t(\ell^{\ast})\sim2\pi $. In the range $\lambda (\ell^{\ast})<\lambda_{c}$
we can find $\lambda (\ell^{\ast})$ by considering the
difference $\lambda_{0}/t_{N}-\lambda_{0}/t_{0}$, using
Eqs. (\ref{pit}) and (\ref{g}). We find
\begin{equation}
t_{N}^{-1}-t_{0}^{-1}\simeq\lambda_{0}^{-1}
\frac{\lambda
(\ell^{\ast})}{3\lambda_{c}}
\bigl (1-\frac{\lambda_{0}}{\lambda_{c}}\bigr )^{1/3},
\end{equation}
from which it follows that
\begin{eqnarray}
\xi \sim e^{\ell^\ast} \sim & & (t_0-t_N(x))^{-1/\epsilon}\times\nonumber\\
& &[1 + Bx(t_0-t_N(x))+...],   
\end{eqnarray}
where $B$ is a constant and
the term associated with it comes from corrections of order 
$\lambda_0 e^{-\ell^\ast}$.
For $\lambda_{0}>t_{0}$, $\lambda (\ell^{\ast})=2\pi $, so that
\begin{equation}
\xi \sim e^{\ell^\ast} \sim (\lambda_{0}^{-1}
 -\lambda_{c}^{-1})^{-1/\epsilon}.
\end{equation}

\begin{figure}
\vspace{1cm}
\centerline{\psfig{figure=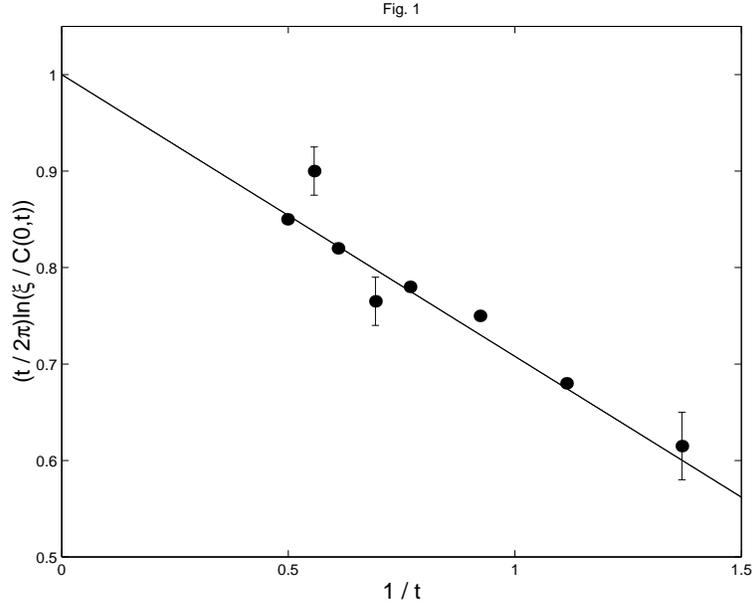,width=10cm}}
\vspace{1cm}
\caption{ $(t/2\pi)\ln(\xi/C)$ versus $1/t$ for La$_2$CuO$_{4+\delta}$ with
$T_N=90$K. The points are from Ref. 4. The straight line shows the fit to
Eq. (54), with $\lambda_0/t_0 \ll 1$.}
\end{figure} 
\begin{figure}
\vspace{1cm}
\centerline{\psfig{figure=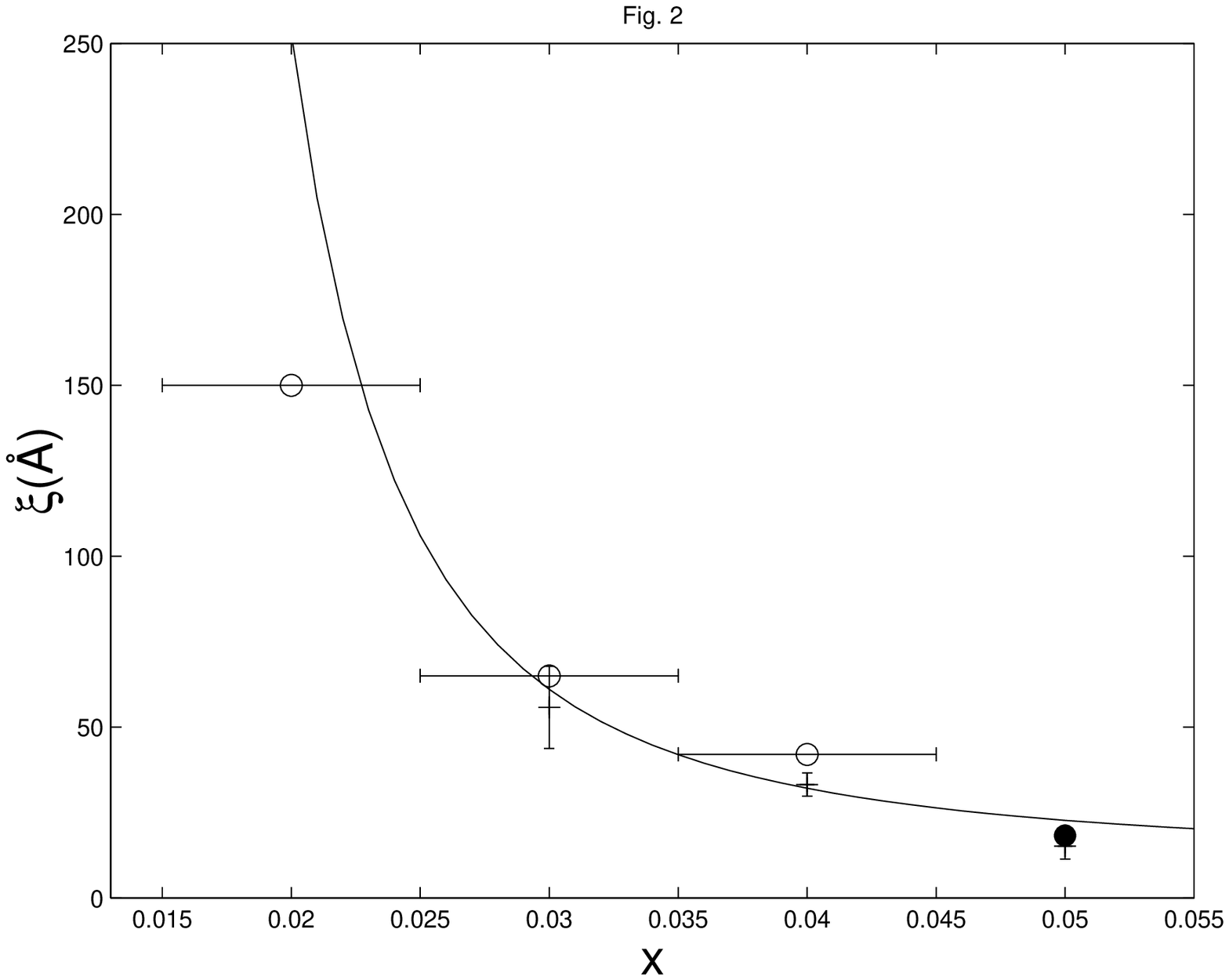,width=10cm}}
\vspace{1cm}
\caption{Dependence of $\xi_{2D}$ at $T=0$ on $x$.  The empty$^4$ and full$^{12}$
 circles indicate
experiments, +'s show numerical simulation$^{27}$ data. The solid line
represents 
$ 2.8 \lambda^{0.8} \AA \exp(2\pi/3\lambda)$, with $\lambda=20x$.}

\end{figure}

\begin{figure}
\vspace{1cm}
\centerline{\psfig{figure=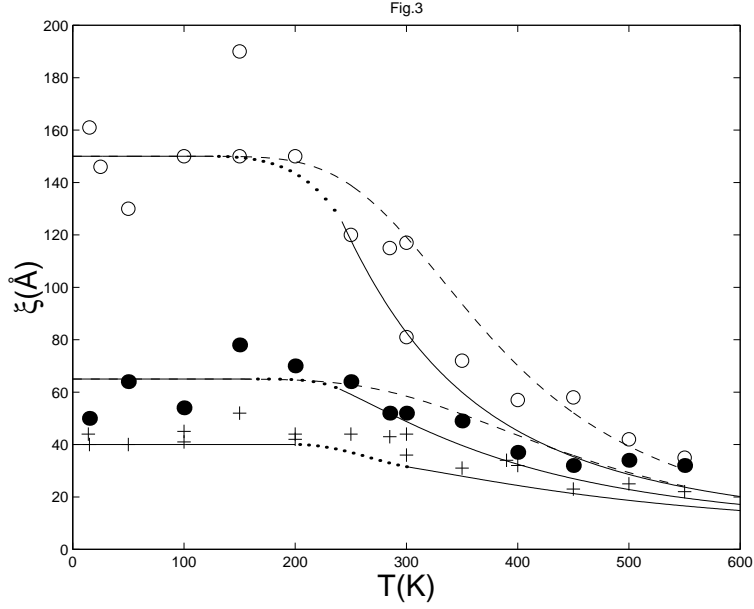,width=10cm}}
\vspace{1cm}
\caption{Dependence of $\xi_{2D}(t,\lambda)$ on $T$ for several concentrations.
Symbols are from experiments: empty circles for $x=0.02$, full circles for
$x=0.03$, +'s for $x=0.04$, all from Ref. 4.
Full lines show results from Fig. 2 (for low $T$)
or Eq. (54), with
$C=L_0\tilde{C}=1.92\AA$ (for high $T$). Dotted lines interpolate between
these low- and high-$T$  theories. Dashed lines correspond to Eq. (69).}
\end{figure}
\begin{figure}
\vspace{1cm}
\centerline{\psfig{figure=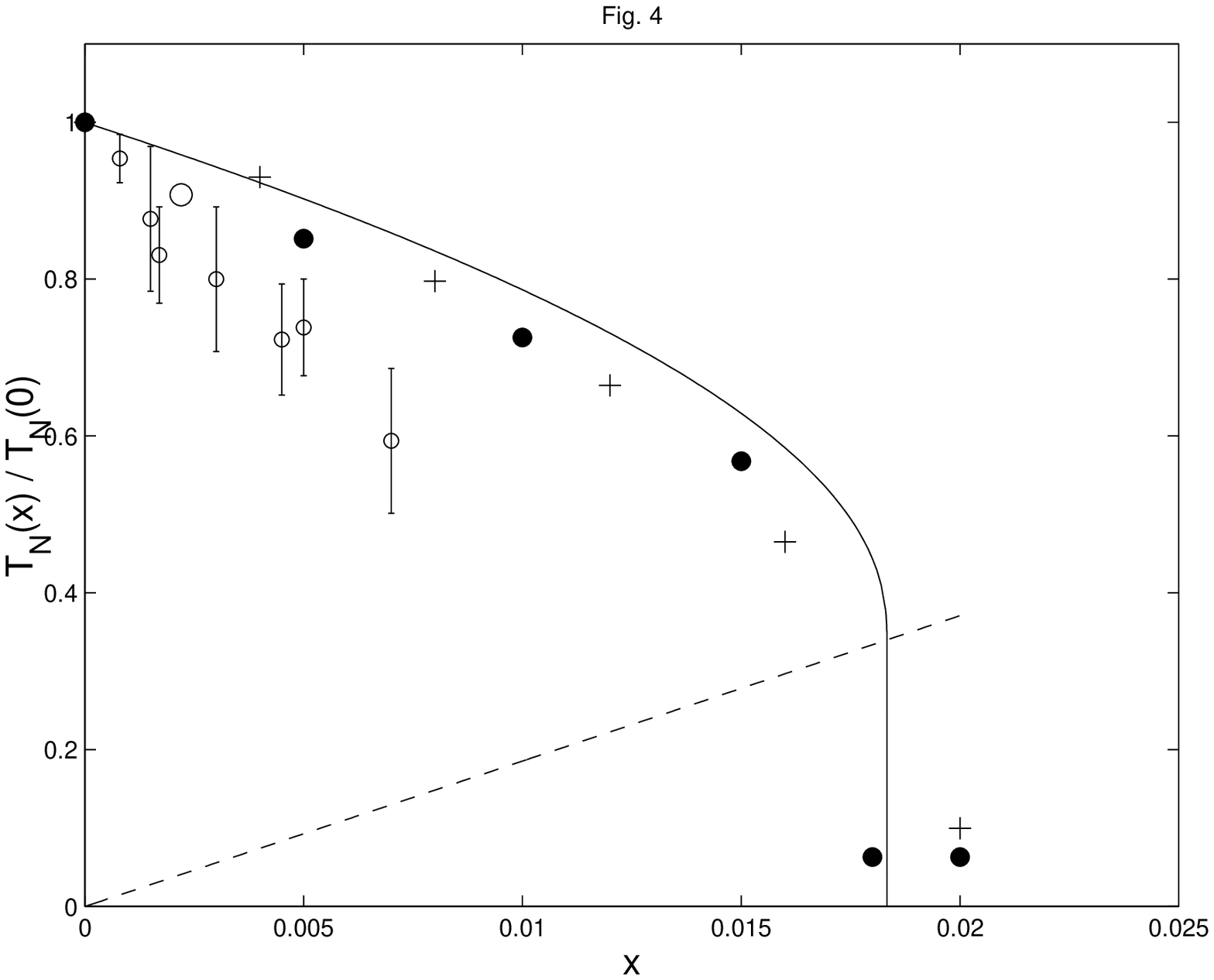,width=10cm}}
\vspace{1cm}
\caption{$T_N(x)/T_N(0)$ versus $x$. Full line is theory, and the points
are from experiments:
 full circles
from Ref.~6 , +'s from Ref.~7, empty circles from Refs. 29 and 31.
Dashed line indicates $\lambda_0=t_0$.}
\end{figure}

%\end{multicols}{2}

\begin{references}
\bibitem{villain} J. Villain, Z. Phys. B{\bf 33}, 31 (1979).
\bibitem{aharony} A. Aharony, R. J. Birgeneau, A. Coniglio,
M. A. Kastner, and H. E. Stanley, Phys. Rev. Lett. {\bf 60}, 1330 (1988).
\bibitem{review} R. J. Birgeneau and G. Shirane, in {\it Physical
Properties of High Temperature Superconductors}, D. M. Ginzberg, ed.,  (World
Scientific, Singapore, 1989).
\bibitem{keimer} B. Keimer, N. Belk, R. J. Birgeneau,
A. Cassanho, C. Y. Chen, M. Greven, M. A. Kastner, A. Aharony, Y. Endoh,
R. W. Erwin, and G. Shirane,
Phys. Rev. B{\bf 46}, 14 034 (1992).
\bibitem{shirane} G. Shirane, R. J. Birgeneau, Y. Endoh, and M. A. Kastner,
 Physica B {\bf 197}, 158 (1994).
\bibitem{saylor} J. Saylor and C. Hohenemser, Phys. Rev. Lett. {\bf 22},
1824 (1990).
\bibitem{cho} J. H. Cho, F. C. Chou, and D. C. Johnston,
 Phys. Rev. Lett. {\bf 70}, 222 (1993).
\bibitem{chou} F. C. Chou, F. Borsa, J. H. Cho, D. C. Johnston, A. Lascialfari,
D. R. Torgeson, and J. Ziolo,
 Phys. Rev. Lett. {\bf 71}, 2323 (1993).
\bibitem{kasegava} H. Kageyama, K. Yoshimura, M. Kato, and K. Kosuge,
 J. Phys. Soc. Jpn. {\bf 64}, 2144 (1995).
\bibitem{wakimoto}S. Wakimoto, K. Kurahashi, C. H. Lee, K. Yamada, Y. Endoh,
and S. Hosoya, Physica B {\bf 237-238}, 91 (1997).
\bibitem{choua} F. C. Chou, N. R. Belk, M. A. Kastner,  R. J. Birgeneau,
and A. Aharony, Phys. Rev. Lett. {\bf 75}, 2204 (1995)
  and references therein.
\bibitem{hayden} S. H. Hayden, G. Appeli, H. Mook, D. Rytz, M. F. Hundley, and
Z. Fisk, Phys. Rev. Lett. {\bf 66}, 821 (1991).
\bibitem{nieder} Ch. Niedermayer, C. Bernhard, T. Blasius, A. Golnik,
A. Moodenbaugh and J. I. Budnick, Phys. Rev. Lett. {\bf 80}, 3843 (1998).
\bibitem{glazman}L. I. Glazman and A. S. Ioselevich, Z. Phys. B{\bf 80},
133 (1990).
\bibitem{korn}
I. Ya. Korenblit, Phys. Rev. B{\bf 51}, 12 551 (1995).
\bibitem{binder}K. Binder and A. P. Young, Rev. Mod. Phys.
{\bf 58}, 80 (1986).
\bibitem{hertz}K. H. Fischer and J. A. Hertz, {\it Spin Glasses} (Cambrige
University Press, 1991).
\bibitem{lacour}P. Lacour-Gayet and G. Toulouse, J. Physique {\bf 35}, 425 (1974)
.
\bibitem{CHN} S. Chakravarty, B. I. Halperin, and D. R. Nelson, Phys. Rev. B{\bf 39}, 2344 (1989).  
\bibitem{imry}Y. Imry and S. K. Ma, Phys. Rev. Lett. {\bf 35}, 1399 (1975).  
\bibitem{brezin}E. Br\'{e}zin and J. Zinn-Justin, Phys. Rev. B{\bf 14}, 
3110 (1976).  
\bibitem{nelson}D. R. Nelson and R. A. Pelcovits, Phys. Rev. B{\bf 16}, 
2191 (1977).  
\bibitem{fisher}D. S. Fisher, Phys. Rev. B{\bf 31}, 7233 (1985).  
\bibitem{polyakov}A. M. Polyakov, Phys. Lett. B{\bf 59}, 79 (1975); 
{\it Gauge Fields and Strings}, in Contemporary Concepts in Physics, 
H. Feshbach, ed., (Harwood Academic, 1987).  
\bibitem{berezinskii}V. L. Berezinskii and A. Ya. Blank, Sov. Phys.  
JETP {\bf 37}, 369 (1973).  
\bibitem{patashinskii}A. Z. Patashinskii and V. L. Pokrovskii,
{\it Fluctuation Theory of Phase Transitions} (Pergamon Press,
Oxford, 1979).
\bibitem{GM}R. J. Gooding and A. Mailhot, Phys. Rev. B{\bf 44}, 11 852 (1991).
\bibitem{HN}P. Hasenfranz and F. Nidermayer, Phys. Lett. B{\bf 268}, 231 (1991).
\bibitem{chen}C. Y. Chen, R. J. Birgeneau, M. A. Kastner, N. W. Preyer,
and T. Thio, Phys. Rev. B{\bf 43}, 392 (1991).
\bibitem{gsbc}R. J. Gooding, N. M. Salem, R. J. Birgeneau, and
F. C. Chou, Phys. Rev. B{\bf 55}, 6360 (1997) and references therein.
\bibitem{chen2} C. Y. Chen, E. C. Branlund, ChinSung Bae, K. Yang,
M. A. Kastner, A. Cassanho, and R.~J.~Birgeneau, Phys. Rev. B{\bf 51},
3671 (1995).
\bibitem{SS} B. Shraiman and E. D. Siggia, Phys. Rev. B{\bf 42}, 2485 (1990).
See also V. L. Pokrovsky and G. V. Uimin, Physica C {\bf 160}, 323 (1989);
R. J. Gooding, Phys. Rev. Lett. {\bf 66}, 2266 (1991);
R. J. Gooding, N. M. Salem and A. Mailhot, Phys. Rev. B{\bf 49}, 6067 (1994)
and Ref. \onlinecite{GM}.
\bibitem{chubukov} A.~V.~Chubukov {\it et al.},  Phys. Rev. B{\bf 49},
11 919 (1994); S.~Sachdev, Phys. Rev. B{\bf 49}, 6770 (1994).
\bibitem{hone} A.~H.~Castro Neto and D.~Hone, Phys. Rev. Lett. {\bf 76},
2165 (1996).


%\bibitem{chubukov} A. V. Chubukov {\it et al.},  Phys. Rev. B{\bf 49},
%11 919 (1994); S. Sachdev, Phys. Rev. B{\bf 49}, 6770 (1994).
\end{references}
\end{document}